\documentclass[12pt, preprint]{aastex}

\usepackage{amsmath,longtable,natbib}
\usepackage{float}
\usepackage{hyperref}
\usepackage{adjustbox}
\usepackage{caption}
\usepackage{graphicx,epstopdf}
\usepackage[usenames,dvipsnames]{color}

\DeclareGraphicsExtensions{.ps}
\shorttitle{}
\shortauthors{Sharma et al.}
\usepackage{subfig}
\begin{document}
\captionsetup[figure]{labelfont={bf},labelformat={simple},labelsep=period,name={Figure}}
\captionsetup[table]{labelfont={bf},labelformat={simple},labelsep=period,name={Table}}
\title{Single-Pulse Correlations in PSR B0329+54: Implications for Radio Emission Zones}
\author{
Shyam~S.~Sharma\altaffilmark{1,2}, Tetsuya Hashimoto\altaffilmark{1}, Tomotsugu Goto\altaffilmark{3}, Sanjay Kudale\altaffilmark{4}, Sujin Eie\altaffilmark{2,5}, Simon C.-C. Ho\altaffilmark{6,7}}

\altaffiltext{1}{Department of Physics, National Chung Hsing University, No. 145, Xingda Rd., South Dist., Taichung 40227, Taiwan}
\altaffiltext{2}{Institute of Astronomy and Astrophysics, Academia Sinica, 11F of AS/NTU Astronomy-Mathematics Building, No.1, Section 4, Roosevelt Road, Taipei 10617, Taiwan}
\altaffiltext{3}{Institute of Astronomy, National Tsing Hua University, Hsinchu 30013, Taiwan}
\altaffiltext{4}{National Centre for Radio Astrophysics, Tata Institute of Fundamental Research, Pune 411007, India}
\altaffiltext{5}{Mizusawa VLBI Observatory, National Astronomical Observatory of Japan, 1-2-2 Mizusawa-Hoshigaoka, Oshu, Iwate 023-0861, Japan}
\altaffiltext{6}{Research School of Astronomy and Astrophysics, The Australian National University, Canberra, ACT 2611, Australia}
\altaffiltext{7}{Centre for Astrophysics and Supercomputing, Swinburne University of Technology, Hawthorn, VIC 3122, Australia}

\section*{ABSTRACT}
\vspace{-3.9mm}

Individual radio pulses from a pulsar are directly linked to the underlying emission processes and the associated magnetic field geometry within its magnetosphere. Thus, single-pulse studies across frequencies can provide crucial insights into the physics of radio emission. {Multiple studies have investigated single-pulse correlations in PSR B0329+54 with widely separated discrete frequencies, reporting the broadband nature of pulsar emission. However, understanding the frequency evolution of these correlations has been limited by poor frequency sampling, and the physical origin of these correlations remains unexplored}. We present a detailed study of single-pulse correlations in PSR~B0329+54 at low radio frequencies using the upgraded Giant Meterwave Radio Telescope (uGMRT), with well-sampled time series spanning 300–1460 MHz. We derived an inverted flux spectrum for this pulsar, with a turnover near 470 MHz. We used flux-calibrated and scintillation-corrected single pulses to study correlations across frequencies. Our results show that maximum correlations consistently occur near the longitude of the central component, with correlation strength exceeding 69\% for all frequency combinations, while outer components exhibit correlations above 46\%. These findings indicate very strong inter-frequency correlations, with no anticorrelations detected. No cross-component correlations were observed; only corresponding components correlate across frequencies. The longitudes of maximum correlation do not coincide with the intensity peaks of the average profile. We also examine how correlations vary with frequency at selected fiducial longitudes. The observations reported in this work favor curvature radiation from relativistic charge bunches in the pulsar plasma; however, reproducing the correlation curves along with spectra remains an open challenge.

\section{Introduction}
\label{sec:intro}
Pulsars emit radio-pulsed emissions at regular intervals of time; a single pulse corresponds to the signal emitted per rotation of the pulsar. In general, the time-averaged profiles of pulsars (averaged emission over hundreds of single pulses) remain quite stable over time (e.g., \citealt{1995ApJ...452..814R}). This stability is particularly intriguing given the highly dramatic nature of single pulses from pulsars (e.g., \citealt{2012MNRAS.423.1351B}). The stability of the averaged profile suggests some underlying, yet not well-understood, systematic behavior of radio emission processes in the pulsar magnetosphere responsible for single-pulse emissions. In the context of an underlying systematic behavior, previous studies (e.g., \citealt{1981A&A....93...85B}, \citealt{1986A&A...163..114K}, \citealt{2003A&A...407..655K}) have reported that single pulse intensity from pulsars exhibits correlations over broad radio frequency observations. In principle, information about the origin of these single pulse emissions, i.e., magnetic field geometry, the size of radio emission zones, and the nature of the emission mechanism, could be explored if careful inspection of these single pulses is carried out over a broad range of frequencies (e.g. \citealt{2024Univ...10..248M}, \citealt{2024ApJ...974..254M}).

In this study, we choose PSR B0329+54 due to its proximity ($\sim$1.64(3) kpc; \citealt{2025PASA...42...98K}) and brightness ($\sim$1.5 Jy at 400 MHz; \citealt{2005AJ....129.1993M}), which make it well-suited for detecting single pulses at low radio frequencies. For PSR B0329+54, \cite{1986A&A...163..114K} reported single-pulse correlations between 102 and 1700 MHz, with a maximum intensity correlation strength of $\sim43\%$ at some pulse phases of the pulsar. One of the key outcomes of their study was that the maximum correlation occurred at unequal pulse phases for the two frequencies. \cite{2003A&A...407..655K} later investigated single-pulse correlations for PSR B0329+54 at 238, 626, 1412, and 4850 MHz, finding a maximum intensity correlation strength of $\sim83\%$ between 626 and 1412 MHz, with values ranging from $39–79\%$ for other frequency combinations. They reported that the single-pulse correlation strength decreases with increasing frequency separation. {Both of these studies make use of sparsely sampled discrete frequencies, which posed a limitation in examining the frequency evolution of the correlation strength. In addition, the origin of these correlations was not explored in either study.}

In this work, we revisit PSR B0329+54 using sensitive single-pulse observations from the upgraded Giant Metrewave Radio Telescope (uGMRT; \citealt{1997hsra.book..217S}, \citealt{2017JAI.....641011R}). While earlier studies examined widely spaced discrete frequencies, we present one of the most densely sampled single-pulse studies of this pulsar at low radio frequencies, covering the 300–1460 MHz range with 13 frequency subbands. Using these data, we seek to address the following three open questions: a) How do the single-pulse correlations evolve at low radio frequencies? b) At what rotational pulse phases are signals at different frequencies most strongly correlated and why? c) Is it possible to uncover the underlying geometry features of the radio emission zones of a pulsar using multi-frequency single-pulse correlations?

Section \ref{sec:Section-2} describes the observations and data reduction, Section \ref{sec:Section-3} describes flux calibration and interstellar scintillation correction in the single-pulse dataset, Section \ref{sec:Section-4} presents the results of the correlation studies, Section \ref{sec:Section-5} discusses possible interpretations, and Section \ref{sec:Section-6} provides a summary of this work.

\section{Observation and Data Reduction}
\label{sec:Section-2}
PSR B0329+54 has DM$=26.74870(7)$ pc$\,$cm$^{-3}$ and a period of $\sim714.5515$ ms, and it has primarily 3 components in the average profile (e.g., \citealt{2025ApJ...980...26S}). We have recorded the total intensity data of PSR B0329+54, as part of proposal DDTC346, using the phased array mode of uGMRT in three bands simultaneously: band-3 (300–500 MHz), band-4 (550–750 MHz), and band-5 (1260–1460 MHz), with a time resolution of 163.84 $\mu$s. The total observation time was $\sim$19 minutes, during which 1594 single pulses were recorded simultaneously at all observed frequencies. We divided each pulse into 4361 equal phase bins, with one bin $\sim$ time resolution. Further details regarding this dataset can be found in Section 2 of \cite{2025ApJ...980...26S}.

\cite{2025ApJ...980...26S} reported that the midpoint of the two outer components in the average profile of PSR B0329+54 closely follows the cold-plasma dispersion law. We carried out the correction for interstellar dispersion delays across frequencies using a DM ($26.7487$ pc$\,$cm$^{-3}$) corresponding to the midpoint of peaks of the two outer components. 

{Furthermore, \cite{2025ApJ...980...26S} estimated that the maximum scattering timescale for PSR B0329+54 within this dataset (300-1460 MHz) is $\sim$ 5 $\mu$s at 300 MHz, based on equations (2) and (3) of \cite{1985ApJ...288..221C} for the Kolmogorov spectrum. Therefore, we ignored the scattering effects in this study, as the predicted maximum scattering timescale ($\sim$ 5 $\mu$s) is insignificant compared to the time resolution (163.84 $\mu$s) of our dataset.}

We divided the frequency range 300-1460 MHz into 13 subbands and averaged each subband to create 13 time series. Table \ref{13_series} lists the frequency intervals and reference frequencies for each.
\begin{table}[H]
    \centering
    \begin{tabular}{|c|c|c|c||c c|}
     \hline
     & \multicolumn{3}{|c||}{Phased Array data} & \multicolumn{2}{|c|}{Interferometric visibility data}\\
     \hline
     uGMRT & Series & Frequency & Reference & B0329+54 & Imaging\\
     
     Band & No. & Range (MHz) &  Frequency (MHz) & Imaging Flux (Jy) & Frequency (MHz)\\ 
     \hline
     \hline
     Band-3 & 1 & 300-317 & 309 & & \\
     & 2 & 317-336 & 327 & 1.4(1) & 334\\
     & 3 & 336-357 & 347 & & \\
     \hline
     & 4 & 357-380 & 369 & 1.27(3) & 375\\
     & 5 & 380-406 & 393 & 1.94(5) & 400\\
     & 6 & 406-434 & 420 & 2.05(2) & 427\\
     & 7 & 434-465 & 450 & 2.1(1) & 456\\
     & 8 & 465-500 & 483 & 2.07(3) & 485\\
     \hline
     \hline
     Band-4 & 9 & 550-590 & 570 & 1.78(5) & 580\\
     & 10 & 590-640 & 615 & 1.42(5) & 635\\
     & 11 & 640-690 & 665 & 1.32(2) & 685\\
     & 12 & 690-750 & 720 & 1.22(6) & 728\\
     \hline
     \hline
     Band-5 & 13 & 1260-1460 & 1374 & 0.14(6) & 1360\\
     \hline
    \end{tabular}
    \caption{Table lists the frequency ranges over which data were averaged to form 13 phased array time series, along with their reference frequencies. It also lists imaging frequencies and flux densities of PSR B0329+54, derived from interferometric visibility data. Flux values used to calibrate the phased array time series are listed alongside each corresponding time series.}
    \label{13_series}
\end{table}

\begin{figure}[H]
    \centering
    \includegraphics[width=1.0\linewidth]{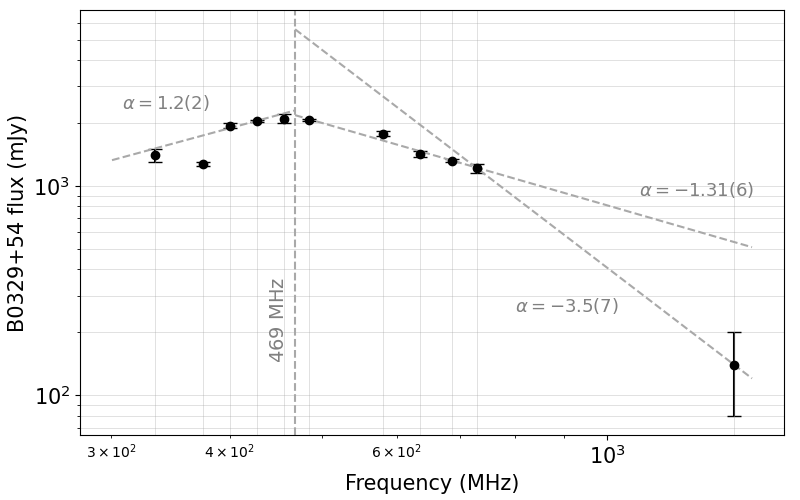}
    \caption{Flux densities of PSR B0329+54 from 300$-$1460 MHz, derived from imaging interferometric visibility data from the uGMRT. The spectrum shows a turnover at 469(8) MHz (dashed vertical line), with flux decreasing on both sides of this turnover frequency. {The flux data above are fitted with a broken power law (dashed rising/declining segments), with $\alpha$ representing the spectral index of the fitted power law.}}
    \label{B0329+54_spectra}
\end{figure}

In parallel to the phased array data, we recorded interferometric visibility data for the same bands simultaneously, with a time resolution of $\sim$2.6 seconds. We carried out a flux calibrator observation (3C147) at the start of the session, and observed a phase calibrator (0432+416) before and after the PSR B0329+54 observation to calibrate the pulsar data. We flagged the data for radio frequency interference (RFI) and calibrated it using the \texttt{FLAGCAL} software \citep{2012ExA....33..157P}; for imaging and self-calibration, we used version 5.6.0-60 of the \texttt{CASA} software\footnote{\url{https://casa.nrao.edu/casadocs/casa-5.6.0}} \citep{2007ASPC..376..127M}. Similar to the phased array data, we initially divided the visibility data into 13 subbands; however, due to high flagging at low radio frequencies, we imaged the visibility data for the first 3 subbands together, resulting in 11 flux values (Table \ref{13_series}). Figure \ref{B0329+54_spectra} shows the flux spectrum of PSR B0329+54 derived from the interferometric visibility observations used in this study. {We followed the approach of modeling the radio flux densities of PSR B0329+54 with broken power laws, as described by \cite{2024ApJ...974..254M}. A five point parabolic fit around the spectral peak gives a turnover frequency of 469(8) MHz. The 334$-$456 MHz flux data can be fitted by a power law with a positive spectral index of 1.2(2) (reduced $\chi^2 \sim 1.9$), excluding the 375 MHz point, which produces an anomalously high $\chi^2$ ($\sim$25). The 485$-$726 MHz flux data can be fitted by a power law with a negative spectral index of –1.31(6) (reduced $\chi^2 \sim 2.7$), while the flux drop between 726 and 1360 MHz requires an additional, steeper power law with an index of –3.5(7).}

\section{Flux Calibration and Correction for Interstellar Scintillation}
\label{sec:Section-3}

Prior to correlating PSR B0329+54 pulses across various frequencies, we flux-calibrated the phased array time series using the imaging continuum flux densities (Table \ref{13_series}) and subsequently corrected for interstellar scintillation. 

\subsection{Flux Calibration}
For flux calibration of time series, we adopted the strategy detailed in \cite{2024ApJ...974..254M}, which is as follows:

1) Remove arbitrary baseline levels from the time series such that the total energy in the pulse $\approx$ ON-pulse energy. Generate the average profile from the baseline-subtracted time series.

2) Calculate the sum of ON-pulse counts, $s_{total}$, in the average profile. Determine the average count, $s_{avg-profile}$, by dividing $s_{total}$ by the number of phase bins, $n_{bins}$, in the average profile, i.e., $s_{avg-profile}=s_{total}/n_{bins}$.

3) Compute the scaling factor by dividing the continuum flux density, $s_{avg-image}$, for that time series (from Table \ref{13_series}) by $s_{avg-profile}$ of the average profile, i.e., scaling factor $=s_{avg-image}$/$s_{avg-profile}$. Multiply the entire time series by this scaling factor.

\subsection{Interstellar Scintillation correction}
Interstellar scintillation modulates the flux of pulsars as a function of time and frequency (e.g., \citealt{1968Natur.218..920S}). Correcting for these effects is essential to determine the intrinsic changes in the pulsar flux densities. 

\begin{figure}[H]
    \centering
    \includegraphics[width=0.9\linewidth]{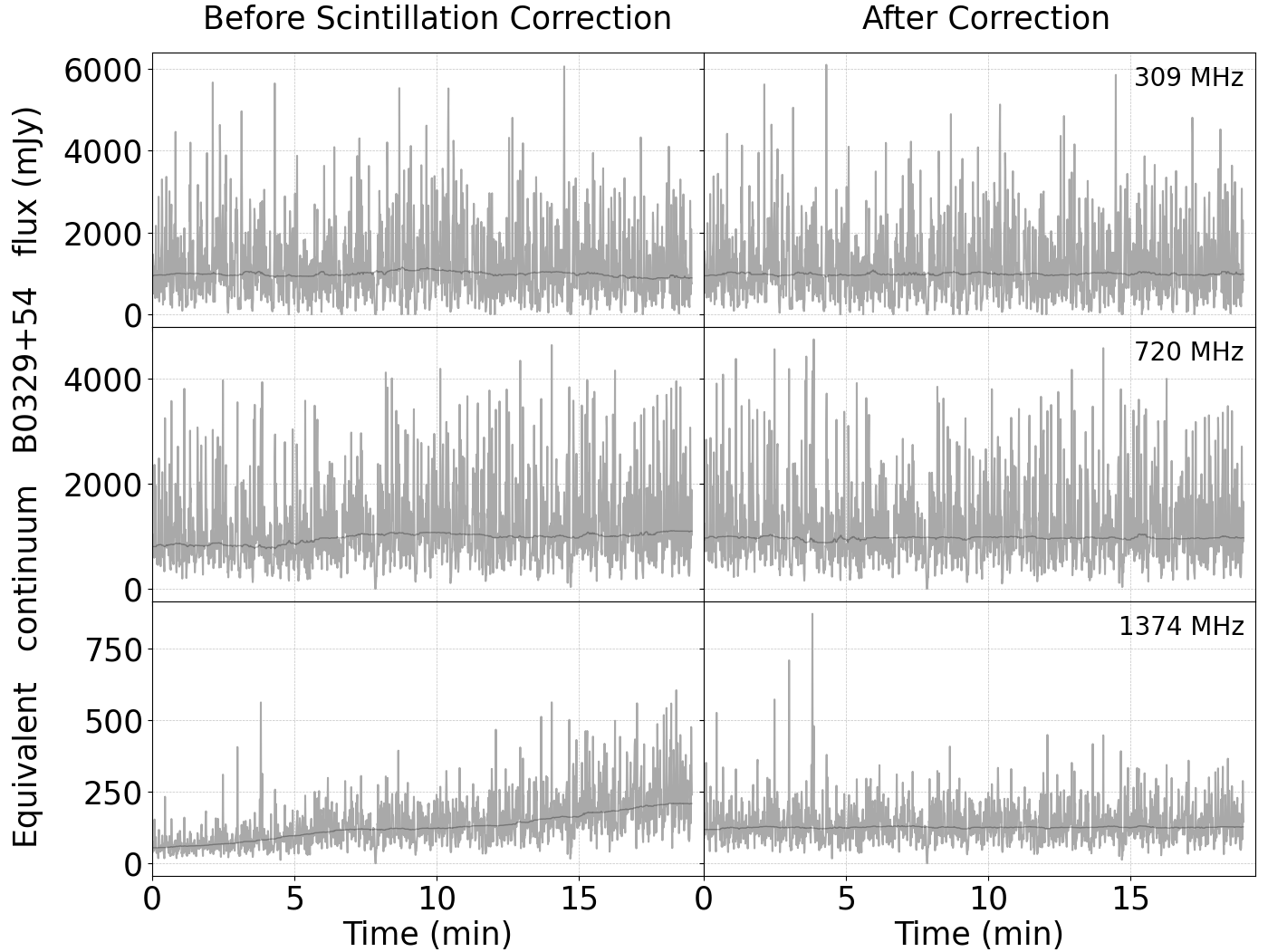}
    \caption{Pulse-to-pulse variations of equivalent continuum flux density for $\sim$19 minutes of observation at 309, 720, and 1374 MHz. The left and right panels display data before and after scintillation correction. The dark solid black line in each plot represents the 200-second running median of the series. In the 309 MHz series, 14 out of 1594 pulses were found to be RFI contaminated, and the corresponding points in this figure have been set to zero.}
    \label{fig_scintillation}
\end{figure}

For frequencies $\sim$200–700 MHz, the predicted diffractive scintillation timescale for PSR B0329+54 ranges from $\sim$2–5 minutes with scintillation bandwidth $<$1 MHz \citep{2003A&A...407..655K}. At $\sim$1.4 GHz, the timescale increases to $\sim$13 minutes with a scintillation bandwidth $\sim$17 MHz \citep{2003A&A...407..655K}. Given our observation duration ($\sim19$ minutes) and the frequency range covered for each time series (Table \ref{13_series}), the scintles should average out for Series 1 (309 MHz) to Series 12 (720 MHz), whereas for Series 13 (1374 MHz), the effects of diffractive scintillation may not be fully averaged out. For this pulsar, \cite{2003A&A...407..655K} calculated the refractive scintillation timescale to be $\sim$830 minutes at $\sim$1.4 GHz, which is significantly longer than both the diffractive scintillation timescales and our observation duration. Figure \ref{fig_scintillation} shows the equivalent continuum flux density ($s_{avg-profile}$ - after flux scaling) per pulse as a function of time for Series 1 (309 MHz), Series 12 (720 MHz), and Series 13 (1.4 GHz), illustrating interstellar flux modulations that may include contributions from both refractive and diffractive scintillation. Similar to the findings of \cite{2003A&A...407..655K}, we observe significant flux changes over time for time series at frequency $\sim$1.4 GHz.

To correct for flux modulations due to interstellar scintillation in PSR B0329+54, we adopt the method described by \cite{2003A&A...407..655K}, which is as follows: 

(1) For each frequency, we first calculate the equivalent continuum flux density ($s_{avg-profile}$) per pulse in the time series, and then compute a 280 pulses ($\sim$ 200 s) running median over this flux series. The 200 s timescale is comparable to or shorter than the diffractive and refractive scintillation timescales at our observing frequencies, and longer than the timescale of intrinsic pulse-to-pulse flux variations of this pulsar. 

(2) Each pulse in the time series is then divided by this running median to remove the interstellar scintillation modulations. 

(3) Finally, the corrected time series is rescaled to match the continuum flux density derived from imaging (Table \ref{13_series}). 
Following this correction, the resulting series (illustrated in Figure \ref{fig_scintillation}) is likely to reflect intrinsic pulse-to-pulse variations originating from the pulsar. The flux calibrated and interstellar scintillation corrected time series are used for further analysis.

\section{Correlation of Single Pulses across Frequency}
\label{sec:Section-4}

In our 13 time series, from 309 MHz to 1374 MHz, each contained a total of M (= 1594) single pulses. We divided each pulse into 4361 equal pulse phase bins (with 1 bin $\sim$ 164 $\mu$s) in each time series. We computed the average profile for each time series by taking the average over 1594 pulses while retaining the resolution of 4361 bins.

We calculated the cross-correlation, \( C_{ij} \), between the pulsar intensities in the \( i \)th rotational phase bin at frequency \( \nu_1 \) and the \( j \)th rotational phase bin at frequency \( \nu_2 \) using the following equation:
\begin{equation}
    C_{ij}=\frac{1}{M}\sum^{M}_{m=1}c_{ij}^m \quad,\quad
    c_{ij}^m=\frac{I_{i}^m[\nu_1] I_j^m[\nu_2] - \bar{I}_i[\nu_1] \bar{I}_j[\nu_2] }{\sigma_{i}[\nu_1]\sigma_{j}[\nu_2]}
    \label{corr_eq}
\end{equation}
\citep{1986A&A...163..114K}. Here, $c^m_{ij}$ corresponds to the correlation between phase bins $i$ and $j$ at frequencies $\nu_1$ and $\nu_2$ for the $m$th single pulse. $I_{k}^m[f]$ is the intensity in the $k$th phase bin of the $m$th pulse of the time series at frequency $f$. $\bar{I}_k[f]$ and $\sigma_{k}[f]$ are the mean and standard deviation of $I_{k}^m[f]$ over $M$ pulses for fixed $k$ and $f$, respectively. The error in $C_{ij}$ is given by $\sqrt{\mathrm{Var}(c_{ij}^m)/M}$, where $\mathrm{Var}(c_{ij}^m)$ is the variance of $c_{ij}^m$ over $M$ pulses.

\subsection{Correlation Maps}
\label{sec:Section-4.1}

\begin{figure}[H]
    \centering
    \includegraphics[width=1.0\linewidth]{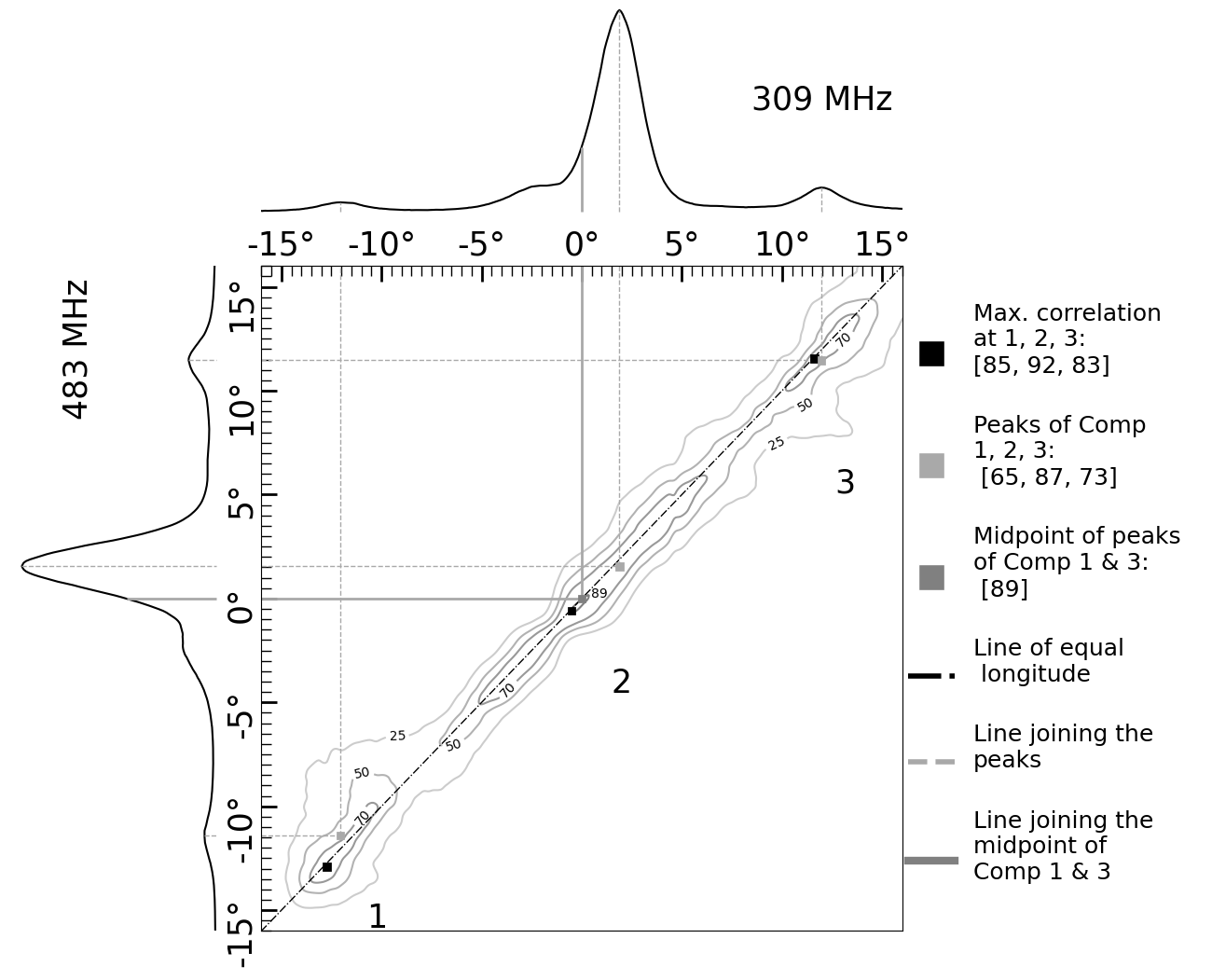}
    \caption{Correlation contour map between the 309 MHz and 483 MHz time series, along with their average profiles. The x- and y-axes represent pulse longitude in degrees.}
    \label{fig_8-1}
\end{figure}

\begin{figure}[H]
    \centering
    \includegraphics[width=1.0\linewidth]{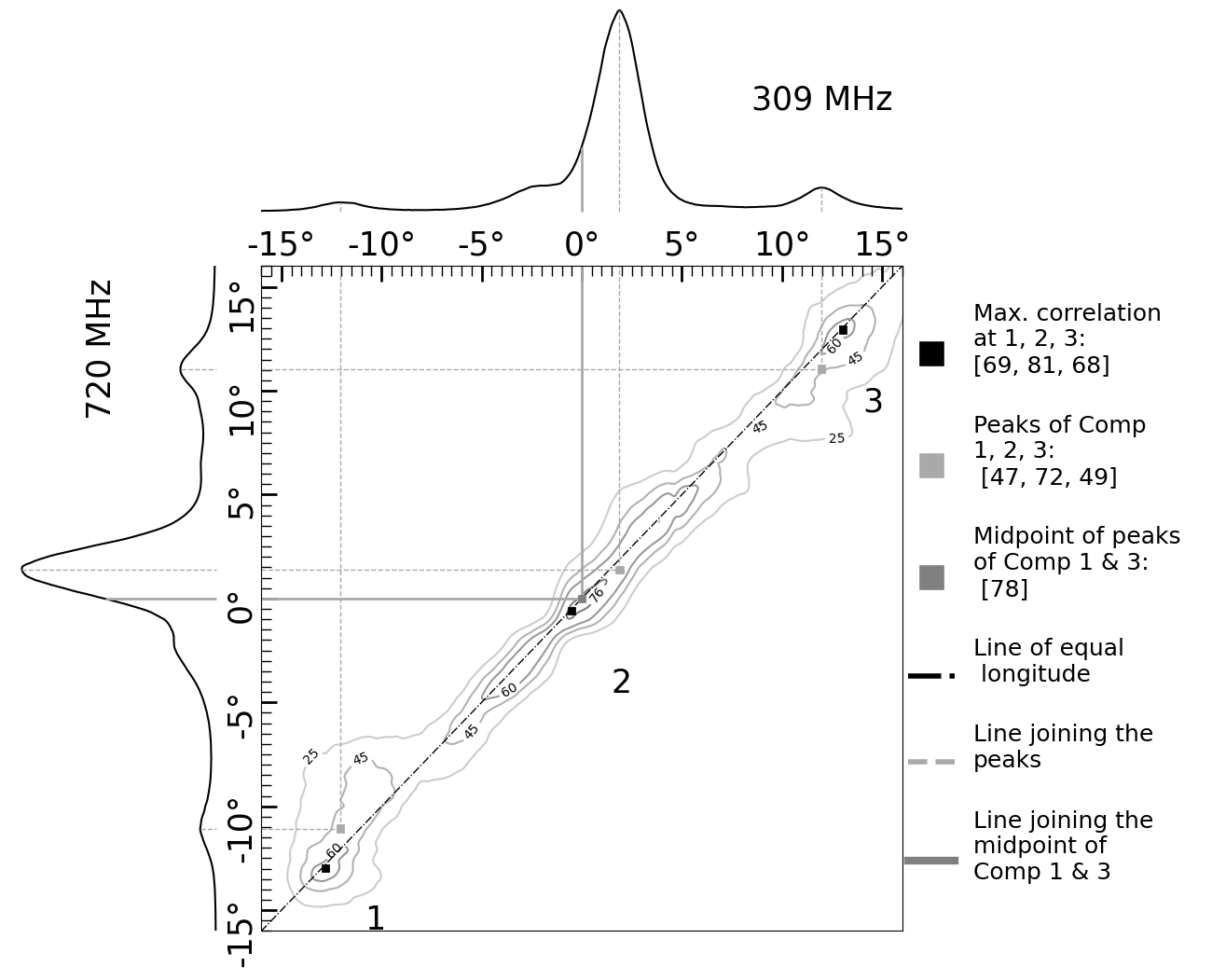}
    \caption{Correlation contour map between the 309 MHz and 720 MHz time series.}
    \label{fig_12-1}
\end{figure}

\begin{figure}[H]
    \centering
    \includegraphics[width=1.0\linewidth]{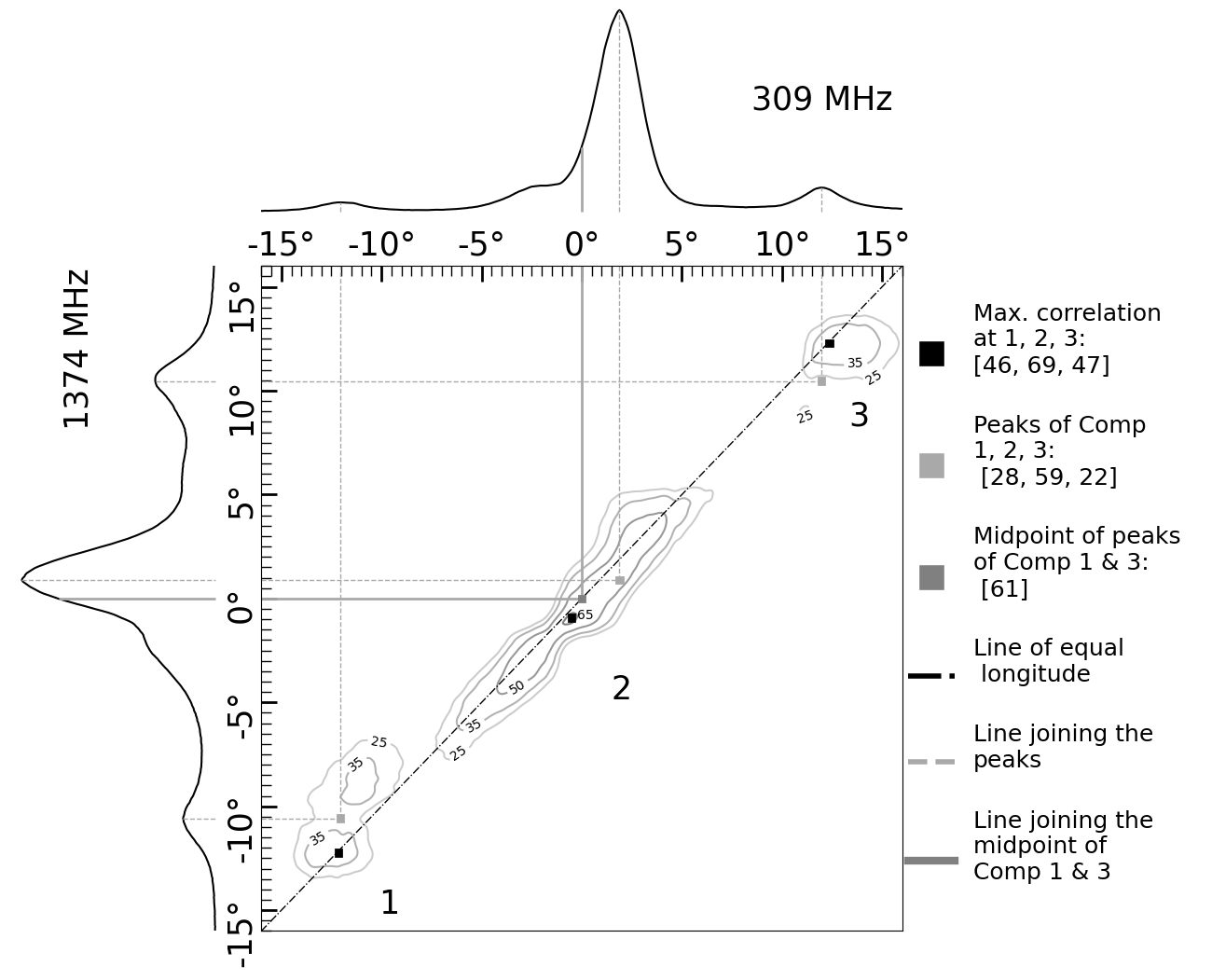}
    \caption{Correlation contour map between the 309 MHz and 1374 MHz time series.}
    \label{fig_13-1}
\end{figure}

In the first attempt, we took the 309 MHz time series as a reference and correlated it with the remaining 12 time series from 327 to 1374 MHz. For illustration purposes, we show the correlation maps for 309 $\&$ 483 MHz, 309 $\&$ 720 MHz, and 309 $\&$ 1374 MHz in Figures \ref{fig_8-1}, \ref{fig_12-1}, and \ref{fig_13-1}, respectively.

We define the three components of PSR B0329+54 in the average profile as Comp 1, Comp 2, and Comp 3. \cite{2011ApJ...741...48C} presented observations of PSR B0329+54 at 610 and 1540 MHz and reported the presence of normal and abnormal emission modes. In the abnormal mode, they found that the trailing component (Comp 3) often becomes less distinguishable from the central component (Comp 2), while the leading component (Comp 1) becomes slightly stronger than Comp 3. More recent studies \cite{2018ApJ...856...55Y} report that the pulsar spends $87\%$ of the time in the normal mode and $13\%$ in the abnormal mode. In addition, \cite{2022MNRAS.512.1906T} report an analysis of a low-emission mode of PSR B0329+54 in which Comp~2 becomes weaker than Comp~1 and Comp~3; this behavior was previously described as “core nulls” in earlier studies (e.g., \citealt{2007MNRAS.379..932M}). \cite{2022MNRAS.512.1906T} further report that this behavior is observed in both the normal and abnormal modes and accounts for $3.6\%$ of the total pulses.

In our 19-minute observation, the three components clearly remained resolved at the observed frequencies, as illustrated by average profiles in Figures \ref{fig_8-1}, \ref{fig_12-1}, and \ref{fig_13-1}, with Comp 3 consistently stronger than Comp 1. This suggests that during our observing session, the pulsar likely remained in its normal emission mode. Therefore, a longer observation duration may enable an investigation of single-pulse correlations across frequencies in both normal, abnormal, and low emission modes offering deeper insights into the pulsar’s emission mechanisms and mode-dependent behavior, motivating future work in this direction.

After the dedispersion process (Section \ref{sec:Section-2}), the midpoint of the peaks of Comp 1 and Comp 3 in the average profile (referred to as the midpoint from now on) of each frequency time series lies at the same pulse phase (or longitude), within $\pm$ 1 time resolution. We have chosen this fixed pulse phase as the origin in all correlation maps.

Each correlation map includes the average profile of the two correlated frequency time series. The contours on the correlation map show the longitudes of equal correlation value. The figures also include a line of equal longitudes (with slope = 1), the points of maximum correlation near the three components, and the points where the intensity peaks of the three components lie in the average profiles. The correlation values at these specific points are given on the right side of the figures.


The correlation values in Equation~\ref{corr_eq} can, in principle, range from –100 to 100. We define a signal as correlated if the correlation value is $\geq 25$, anti-correlated if $\leq -25$, and uncorrelated if it lies between –25 and 25.

All values in the correlation maps remain above –16, i.e., no anticorrelation is observed between signals at different frequencies. The correlated signal consistently appears along a straight line, and no cross-component correlation is present between the two frequencies; only corresponding components correlate with each other. 

Another observation from the figures is that the longitude of maximum correlation in the correlation maps lies in the vicinity of the midpoint. The correlation value at the midpoint always remains $\geq61 \%$ for all frequency correlation maps, indicating a very strong correlation across frequencies at this location. The correlation values at the midpoint are within $\sim 12 \%$ of the highest correlation observed for all frequency maps.

The peak of Comp 1 in the average profile, located on the left side of the midpoint, shows a continuous upward shift in the correlation maps from 309–483 MHz to 309–720 MHz and further to 309–1374 MHz. This shows that Comp 1 moves closer to the midpoint as frequency increases. On the right side of the midpoint, the peaks of Comp 2 and Comp 3 show a continuous downward shift in the correlation maps with increasing frequency, likewise showing a shift of components toward the midpoint. This behavior is also reflected in the correlation contours, where the correlation island near Comp 1 shows an upward shift relative to the lines of equal longitude as the frequency increases, the island near Comp 3 appears to move downward, and the island corresponding to Comp 2 shows a subtle downward shift in its weight distribution (considering contours of high correlations). From both the single-pulse correlations and the average profile peaks, it is observed that all three components move closer to the midpoint as the frequency increases from 309 to 1374 MHz. 

The longitudes of maximum correlation near the three components of the average profile are not aligned with the longitudes of the intensity peaks themselves. For example, the point of maximum correlation near Comp 2 lies close to the midpoint but is located farther in longitude from the intensity peak of Comp 2. It shows that for PSR B0329+54, the intensity peaks in the average profile do not correspond to the highest correlated points across frequencies at the single-pulse level.

\subsection{Variation in Longitudes of Correlation Maxima}
\label{sec:Section-4.2}

\begin{figure}[H]
    \centering    \includegraphics[width=1.0\linewidth]{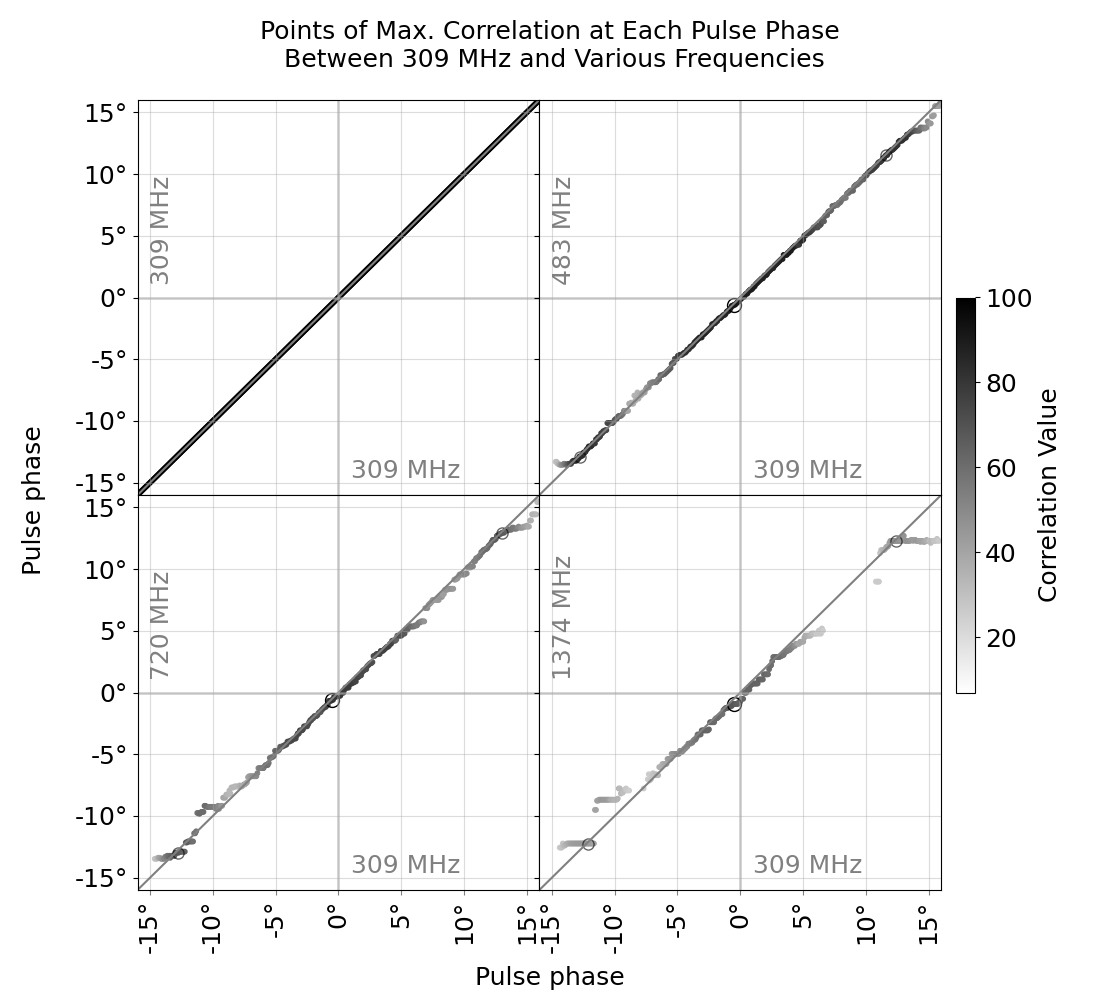}
    \caption{Maximum correlation points at each pulse longitude of 309 MHz signal when correlated with various frequencies.}
    \label{max_correlation_309}
\end{figure}

Figure \ref{max_correlation_309} shows the points of maximum correlation at each longitude of the 309 MHz signal when correlated with various frequency signals (309, 483, 720, and 1374 MHz), indicating which longitude of the other frequency (e.g., 483 MHz) has the strongest correlation with a given longitude of the 309 MHz. The x- and y-axes for each panel in Figure \ref{max_correlation_309} are the same as those for Figures \ref{fig_8-1}, \ref{fig_12-1}, and \ref{fig_13-1}. For reference, each correlation panel includes a line of equal longitude (with slope=1).

For instance, when the 309 MHz signal is correlated with itself, the points of maximum correlation align along the line of equal longitudes, all having a value of 100 (top-left panel of Figure \ref{max_correlation_309}). The open circles in the other three panels mark the locations of maximum correlation near the three components of the average profile of PSR B0329+54.

In Figure \ref{max_correlation_309}, as the frequency separation increases, the points of maximum correlation progressively deviate from the line of equal longitudes. Specifically, the points on the left side of the midpoint shift upward with increasing frequency, while those on the right side shift downward. The points located just to the right of the midpoint exhibit a subtle downward displacement. This confirms the movement of components toward the midpoint as the frequency increases, consistent with the results discussed in Section \ref{sec:Section-4.1}. 

In addition, a new feature stands out in Figure \ref{max_correlation_309}. Specifically, small horizontal flat patches are observed in the 309–1374 MHz panel, showing that a longitudinal range of the 309 MHz signal is maximally correlated with a single longitude of the 1374 MHz signal. This effect is more pronounced for the outer components of PSR B0329+54. For small horizontal flat patches, the correlation is highest near the equal-longitude line in 309-1374 MHz panel and diminishes with increasing pulse-phase separation between them. For the central component, the points of maximum correlation at each longitude largely follow the line of equal longitudes, with only minor deviations.

\subsection{Correlation at Specific Longitudes; 309 MHz as Reference}
\label{sec:Section-4.3}

\begin{figure}[H]
    \centering    \includegraphics[width=1.0\linewidth]{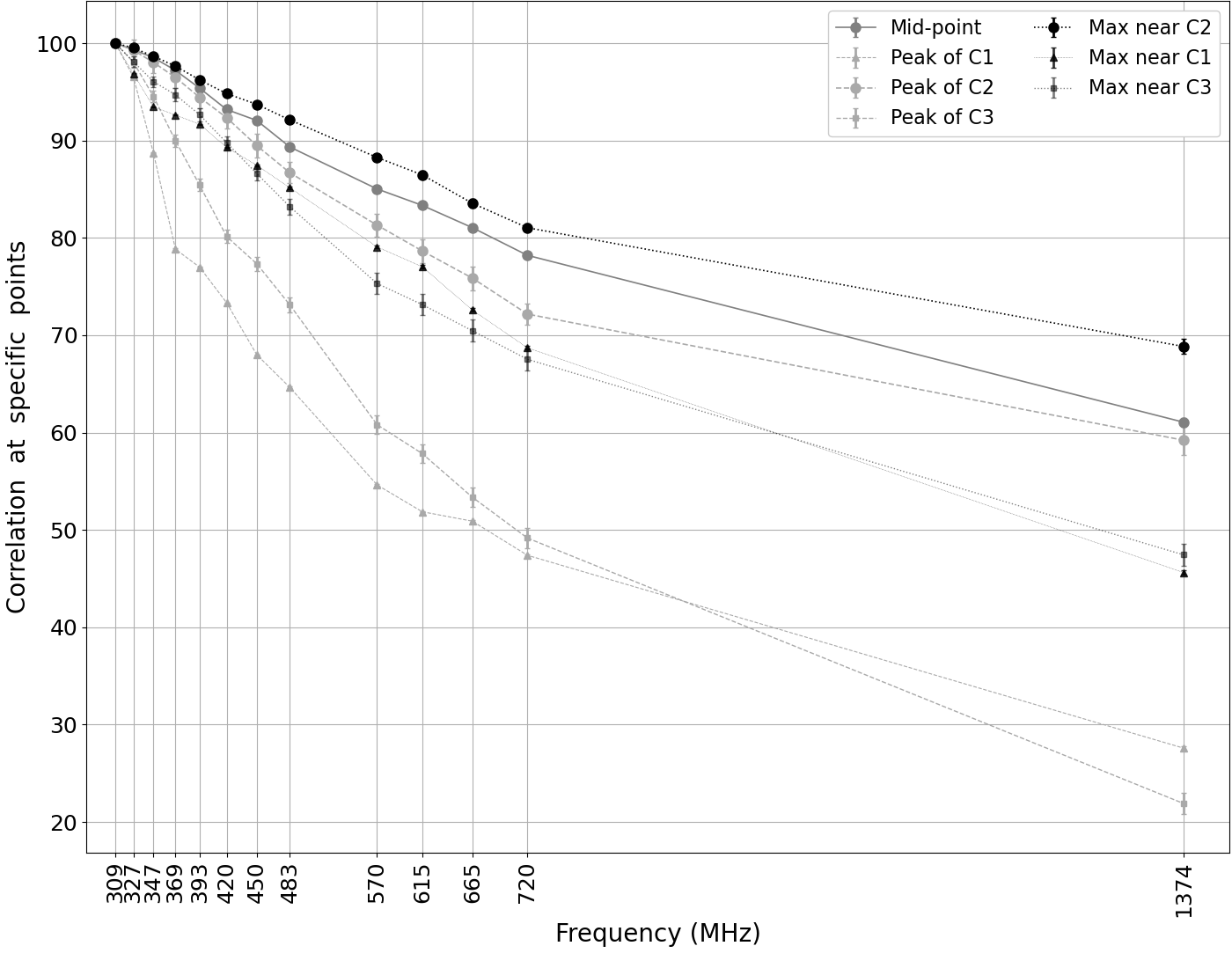}
    \caption{Correlation values at specific marker locations when the 309 MHz signal is correlated with various frequencies.}
    \label{7_point_correlation}
\end{figure}

Figure \ref{7_point_correlation} shows the correlation values at seven specific points: (i) the midpoint, (ii) the intensity peaks of the three components in the average profile, and (iii) the maximum correlation points near the three components. These seven values are extracted from each of the 13 correlation maps, which are generated by correlating the 13 time series with a common 309 MHz time series.

The maximum correlation consistently occurs near Comp 2, with the midpoint often showing a value closest to this maximum. The intensity peak of Comp 2 shows a lower correlation value than both the midpoint and the maximum correlation point.

Figure \ref{7_point_correlation} shows that although these seven points remain correlated across all frequency pairs in general, the correlation values systematically decrease with increasing frequency separation from 309 MHz. Also, the correlation points, which are nearby to Comp 2 (see top three curves), show a lesser decrease in correlation as the frequency separation increases from 309 MHz compared to the points near to Comp 1 and Comp 3 (see bottom four curves). In addition, for all seven curves, the correlation is observed to decrease more rapidly at lower frequencies compared to higher frequencies.

\subsection{Correlation at midpoint; Varying Reference Frequencies}
\label{sec:Section-4.4}

The midpoint longitude remains fixed across all frequencies, while the longitudes of the other six points vary with frequency. It is of interest to examine how the correlation behaves at the midpoint when different reference frequencies (other than 309 MHz) are used for the analysis.

\begin{figure}[H]
    \centering    \includegraphics[width=1.0\linewidth]{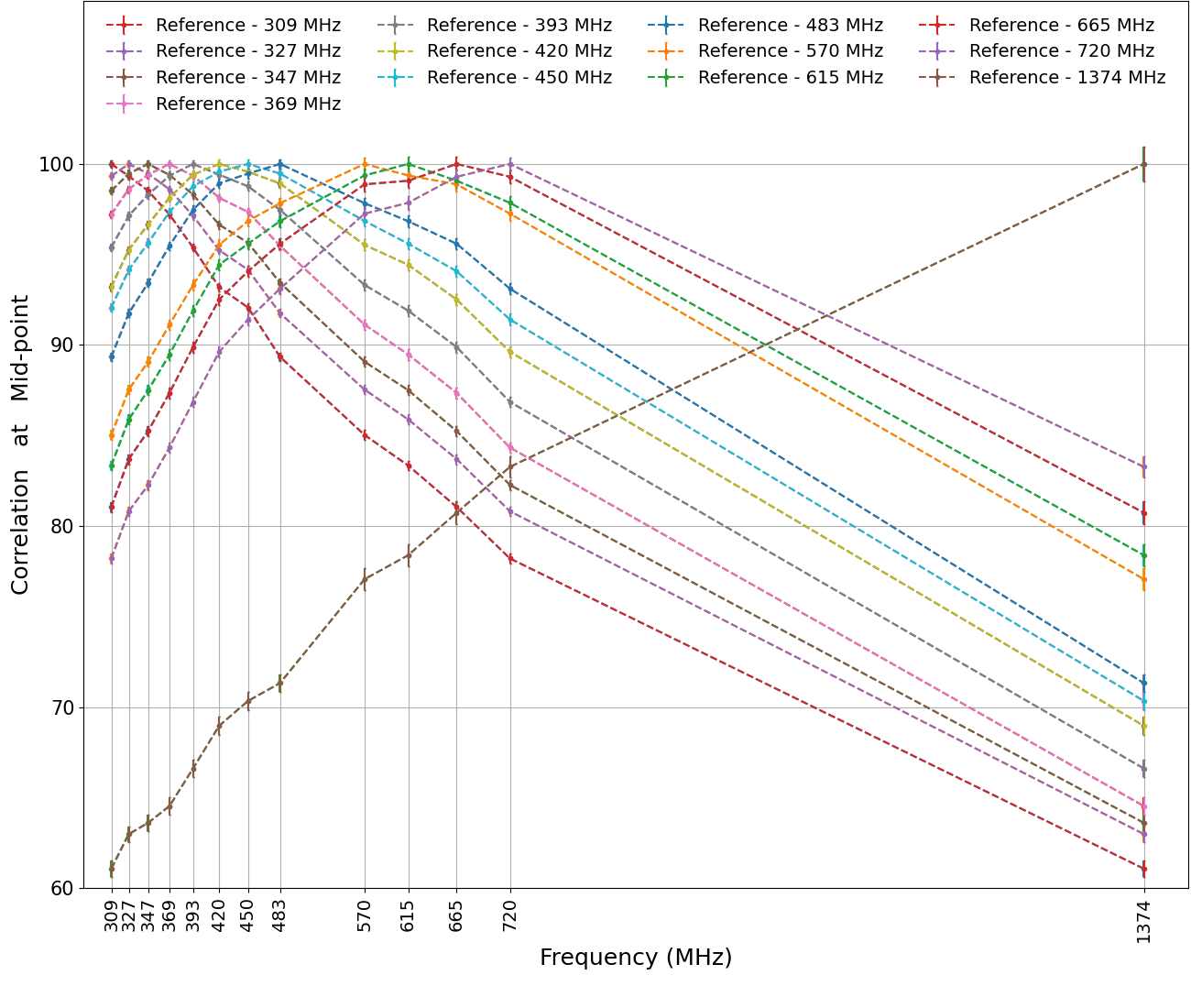}
    \caption{Correlation value at the midpoint when each time series is used as the reference and correlated with various frequencies.}
    \label{mid_point_correlation}
\end{figure}

Figure \ref{mid_point_correlation} shows 13 correlation curves corresponding to the midpoint, obtained by using each of the 13 frequency time series as the reference and correlating it with all 13 frequencies time series one by one. For example, the first (decreasing) curve from left is obtained when the 309 MHz time series is used as a reference and correlated with 13 time series, resulting in 13 correlation values at the midpoint as a function of frequency. Similarly, the second curve from left corresponds to using 327 MHz as the reference, and so on. Note that when a frequency signal is correlated with itself, the correlation at the midpoint reaches $100\%$.

In Figure \ref{mid_point_correlation}, for all 13 curves, the correlation at the midpoint decreases as the frequency separation between the two signals increases. Interestingly, for the curves where a decrease in correlation is seen on both sides of the reference frequency, the decline is observed to be steeper toward lower frequencies and comparatively less steep toward higher frequencies. That is, for a given reference frequency $f_0$, the correlation with $f_0 + \Delta f$ is consistently higher than with $f_0 - \Delta f$. For instance, this observation is clearly illustrated by the curve corresponding to 483 MHz as the reference (the eighth curve from the left). This result indicates an apparent faster decorrelation of signals at lower frequencies compared to higher frequencies.

\section{Discussion}
\label{sec:Section-5}
The coherent curvature radiation in pulsar plasma is considered the primary source of observed radio emission in pulsars (e.g., \citealt{2023MNRAS.521L..34M}, \citealt{1999ptep.proc..381M}, \citealt{1975ApJ...196...51R}, \citealt{1969ApJ...157..869G}). For PSR B0329+54, \cite{2024Univ...10..248M}  reported a successful fitting of the rotating vector model (\citealt{1969ApL.....3..225R}) in polarization position angle data, which naturally favors coherent curvature radiation as the emission mechanism. The standard model is as follows: In the pulsar plasma, relativistic charge bunches accelerate along open dipolar magnetic field lines and produce radio emission. Higher frequencies originate at lower altitudes, while lower frequencies arise at higher altitudes. Recently, \cite{2025ApJ...985..247B} conducted detailed numerical simulations of emission from charge bunches in pulsar plasma, using dipolar magnetic field geometry and curvature radiation as the emission mechanism. They showed that the higher radio frequencies originate within a narrow altitude range along open magnetic field lines, while lower frequencies arise from a broader altitude range, as demonstrated in Figure 6 of \cite{2025ApJ...985..247B}.

\begin{figure}[H]
    \centering    \includegraphics[width=0.8\linewidth]{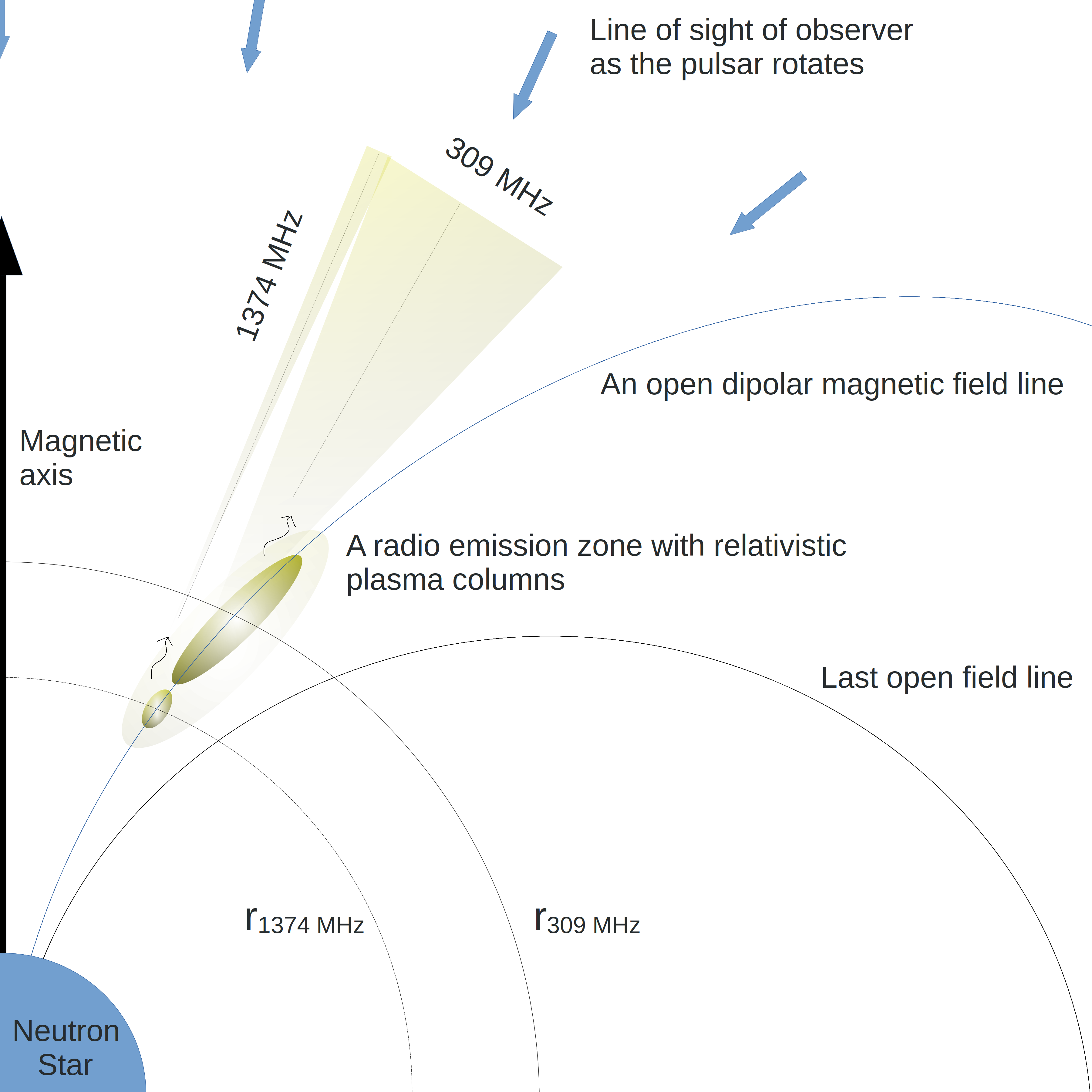}
    \caption{Schematic diagram of PSR B0329+54 illustrating radio emission along an open dipolar magnetic field line.}
    \label{schematic}
\end{figure}

Multiple studies (e.g., \citealt{1991A&A...243..219G}, \citealt{2025ApJ...980...26S}) proposed geometrical models for PSR B0329+54. Based on these models and the curvature radiation characteristics from charge bunches, described by \cite{2025ApJ...985..247B}, we constructed a schematic diagram (Figure \ref{schematic}) to interpret the observations in this work. The figure illustrates one radiation zone along an open dipolar field line responsible for a single observed emission component of the pulsar. The observer's line of sight lies in the plane of the magnetic axis and the emission corresponding field line, with the midpoint located along the magnetic axis. The 309~MHz emission peaks at altitude $r_{309\,\texttt{MHz}}$ above the star, and the 1374~MHz emission peaks at $r_{1374\,\texttt{MHz}}$, where $r_{309\,\texttt{MHz}} > r_{1374\,\texttt{MHz}}$.

In section \ref{sec:Section-4.1}, we find no cross-component correlation between various frequencies; only corresponding components correlate with each other. This suggests that different components originate from spatially separated set of field lines along which charge bunches can emit independently, consistent with the geometrical model shown in Figure 6 of \cite{2025ApJ...980...26S}.

In Sections \ref{sec:Section-4.1} and \ref{sec:Section-4.2}, we observed that the emission components move closer to the midpoint (located along the magnetic axis) as the frequency increases. This behavior arises naturally in dipolar magnetic field geometry, where components move closer to magnetic axis at lower altitudes (high-frequency peaking regions) and farther apart at higher altitudes (low-frequency peaking regions).

The charge bunches move at relativistic speeds, close to the speed of light, along the field lines. The field lines where the motion of charge bunches aligns with the direction of propagation of the radiation will be observed simultaneously. Along such field lines, the same charge bunches emit at different altitudes, producing highly correlated radiation at different radio frequencies. This may explain the higher correlation among frequencies near the central components observed in Sections \ref{sec:Section-4.1}, \ref{sec:Section-4.2}, and \ref{sec:Section-4.3}. For the outer components, the faster decrease in correlation arises similarly: {magnetic field lines curve more if they are farther from the magnetic axis}. The field-line flaring increasingly separates the motion of charge bunches from the radiation direction at higher altitudes, causing a faster drop in correlation compared to the central components. 

In Section \ref{sec:Section-4.2}, we observe that a longitudinal range at 309 MHz correlates with a single longitude at 1374 MHz (Figure \ref{max_correlation_309}). This may be explained by the simulated emission pattern from charge bunches reported by \cite{2025ApJ...985..247B}, where lower radio frequencies originate from a wide altitude range and higher radio frequencies from comparatively narrower altitude range along a field line (see Figure \ref{schematic}).

In Sections \ref{sec:Section-4.3} and \ref{sec:Section-4.4}, it has been observed that correlations between two signals decrease as their frequency separation increases. This may be explained if lower frequency radiation being emitted at increasingly higher altitudes, leading to reduced correlation. We also observed that correlation between two frequency signals decreases faster at lower frequencies than at higher frequencies for the same frequency difference. This may suggest that the altitude separation between two signals is larger at lower frequencies than at higher frequencies for the same frequency interval.

In this work, many of the observed emission features may be readily interpreted using the dipolar magnetic field geometry and the curvature radiation emission mechanism. However, reproducing the correlation curves in Figure \ref{mid_point_correlation} remains a challenge and requires further numerical simulations, similar to the work of \cite{2025ApJ...985..247B}. Identifying charge structures that could simultaneously produce these correlation curves and the inverted spectrum of PSR B0329+54, with a turnover frequency around 470 MHz (Figure \ref{B0329+54_spectra}), represents a potential direction for future investigations.

\section{Summary}
\label{sec:Section-6}
We reported the best-sampled single-pulse correlation investigations to date at low radio frequencies for PSR B0329+54 using the uGMRT, utilizing 13 distinct frequency time series from 300 to 1460 MHz. In parallel to single-pulse time series, we recorded continuum imaging interferometric data, from which we derived the spectrum of this pulsar across 300-1460 MHz, with an inverted shape and turnover frequency $\sim$ 470 MHz. 

We calibrated the single-pulse dataset with the corresponding continuum imaging fluxes, following the technique described by \cite{2024ApJ...974..254M}.

We revisited the interstellar scintillation correction technique described by \cite{2003A&A...407..655K} and applied it to our single-pulse time series. We found that the correction is only necessary for the 1374 MHz time series, while those below 750 MHz show no substantial interstellar modulation.

The flux-calibrated and scintillation-corrected single-pulse dataset was used to investigate correlations among different frequency signals. The findings are as follows:

1) We found no anticorrelation between various frequency signals for this pulsar. No cross-component correlations were detected; correlations occur only between corresponding components across frequencies.

2) The emission components shift toward the midpoint with increasing frequency.

3) The longitudes of maximum correlation between two frequency signals do not coincide with the intensity peaks of components in the average profile.

4) The longitude of the midpoint lies in the vicinity of profile's maximum correlation point and remains highly correlated ($\geq$ 61$\%$) between different frequency signals across 300-1460 MHz. The maximum correlation near the central component is always $\geq$ 69$\%$, and near the outer components, it is $\geq$ 46$\%$ across all frequency combinations in our data.

5) The correlation between two frequency signals is lower for the outer components than for the central component.

6) For the outer components, a longitudinal range of 309 MHz is correlated with a single-pulse phase at 1374 MHz, and this is more pronounced for the outer components. Whenever this effect is observed, the correlation is maximum near equal longitudes of the two frequency signals and gradually decreases as the separation between their pulse phases increases. 

7) The correlation between two frequency signals decreases as the frequency separation increases.

8) The correlation between two frequency signals is lower at lower radio frequencies than at higher frequencies for the same frequency difference of two frequencies.
 
9) With the midpoint aligned along the magnetic axis and the observer’s line of sight passing close to it as the pulsar rotates, the observations can be readily interpreted using the following ideas:
 
i) Emission for different components originates from distinct sets of field lines, which are spatially separated and act as independent emission zones.

ii) The altitudes of radio emission differ between frequencies, with lower frequencies emitted at higher altitudes and higher frequencies from lower altitudes above the star.

iii) The altitude difference between two signals is greater at lower radio frequencies than at higher frequencies for the same frequency interval.

iv) The charge bunches responsible for radio emission move at relativistic speeds, close to the speed of light, along open dipolar field lines. The field lines flare more at higher altitudes as the radio frequency decreases.

v) Lower radio frequencies originate from a wider altitude range compared to higher frequencies.

{This study provides a detailed analysis of the correlations for PSR B0329+54 at low radio frequencies. It also qualitatively bridges the gap between previously reported theoretical simulations of emission features from charge bunches and the observed correlations among frequencies in this work. In addition, this work demonstrates how the magnetospheric geometry responsible for emission imprints can be inferred from single-pulse correlation studies}.

Many of the observations in this work may be readily connected with the theoretical models of curvature radiation along open dipolar field lines. However, reproducing the correlation curves together with the inverted spectra of this pulsar remains challenging, as it requires a precise determination of the charge structures, their speeds along the field lines, and the plasma properties in the magnetosphere responsible for the radio emission. Simulating the observations reported here provides a promising direction for future studies.

\section{Acknowledgments}
We would like to thank Professor \textbf{Dipanjan Mitra} (NCRA-TIFR), who helped to take the suitable type of observations from GMRT, which can address the key questions regarding the physics of radio emission in pulsars. He also provides guidance from time to time during the course of this work. SSS acknowledges the support of the National Science and Technology Council of Taiwan through grants 113-2123-M-001-008- and 114-2811-M-005 -015 -MY2.
TH is very grateful to the Ministry of Science and Technology of Taiwan through grants 113-2112-M-005-009-MY3, 113-2123-M-001-008-, 111-2112-M-005-018-MY3, and the Ministry of Education of Taiwan through a grant 113RD109. We thank the anonymous reviewer for their comments on this work.


\begin{thebibliography}{}
\bibitem[Kumar et al.(2025)]{2025PASA...42...98K} Kumar, A., Deller, A.~T., Jain, P., et al.\ 2025, pasa, 42, e098. doi:10.1017/pasa.2025.10063

\bibitem[Bartel et al.(1981)]{1981A&A....93...85B} Bartel, N., Kardashev, N.~S., Kuzmin, A.~D., et al.\ 1981, \aap, Simultaneous two-station single pulse observations of radio pulsars over a broad frequency range., 93, 85. 

\bibitem[Basu et al.(2025)]{2025ApJ...985..247B} Basu, R., Mitra, D., Melikidze, G.~I., et al.\ 2025, \apj, 985, 2, 247. doi:10.3847/1538-4357/adce7a

\bibitem[Burke-Spolaor et al.(2012)]{2012MNRAS.423.1351B} Burke-Spolaor, S., Johnston, S., Bailes, M., et al.\ 2012, \mnras, The High Time Resolution Universe Pulsar Survey - V. Single-pulse energetics and modulation properties of 315 pulsars, 423, 2, 1351. doi:10.1111/j.1365-2966.2012.20998.x

\bibitem[Chen et al.(2011)]{2011ApJ...741...48C} Chen, J.~L., Wang, H.~G., Wang, N., et al.\ 2011, \apj, 741, 1, 48. doi:10.1088/0004-637X/741/1/48

\bibitem[Cordes et al.(1985)]{1985ApJ...288..221C} Cordes, J.~M., Weisberg, J.~M., \& Boriakoff, V.\ 1985, \apj, 288, 221. doi:10.1086/162784

\bibitem[Gil(1991)]{1991A&A...243..219G} Gil, J.~A.\ 1991, \aap, 243, 219. 

\bibitem[Goldreich \& Julian(1969)]{1969ApJ...157..869G} Goldreich, P. \& Julian, W.~H.\ 1969, \apj, 157, 869. doi:10.1086/150119

\bibitem[Kardashev et al.(1986)]{1986A&A...163..114K} Kardashev, N.~S., Nikolaev, N.~Y., Novikov, A.~Y., et al.\ 1986, \aap, 163, 114

\bibitem[Kramer et al.(2003)]{2003A&A...407..655K} Kramer, M., Karastergiou, A., Gupta, Y., et al.\ 2003, \aap, Simultaneous single-pulse observations of radio pulsars. IV. Flux density spectra of individual pulses, 407, 655. doi:10.1051/0004-6361:20030842

\bibitem[Manchester et al.(2005)]{2005AJ....129.1993M} Manchester, R.~N., Hobbs, G.~B., Teoh, A., et al.\ 2005, \aj, 129, 1993. doi:10.1086/428488

\bibitem[McMullin et al.(2007)]{2007ASPC..376..127M} McMullin, J.~P., Waters, B., Schiebel, D., et al.\ 2007, Astronomical Data Analysis Software and Systems XVI, 376, 127

\bibitem[Melikidze et al.(1999)]{1999ptep.proc..381M} Melikidze, G.~I., Gil, J.~A., \& Pataraya, A.~D.\ 1999, Plasma Turbulence and Energetic Particles in Astrophysics, 381.

\bibitem[Mitra et al.(2007)]{2007MNRAS.379..932M} Mitra, D., Rankin, J.~M., \& Gupta, Y.\ 2007, \mnras, 379, 3, 932. doi:10.1111/j.1365-2966.2007.11988.x

\bibitem[Mitra et al.(2023)]{2023MNRAS.521L..34M} Mitra, D., Melikidze, G.~I., \& Basu, R.\ 2023, \mnras, 521, 1, L34. doi:10.1093/mnrasl/slad022

\bibitem[Mitra et al.(2024a)]{2024Univ...10..248M} Mitra, D., Basu, R., \& Melikizde, G.~I.\ 2024, Universe, Decoding the Nature of Coherent Radio Emission in Pulsars I: Observational Constraints, 10, 6, 248. doi:10.3390/universe10060248

\bibitem[Mitra et al.(2024b)]{2024ApJ...974..254M} Mitra, D., Basu, R., \& Melikidze, G.~I.\ 2024, \apj, On the Flux Density Spectral Property of High Linearly Polarized Signal from Pulsar J0332+5434, 974, 2, 254. doi:10.3847/1538-4357/ad71ca

\bibitem[Prasad \& Chengalur(2012)]{2012ExA....33..157P} Prasad, J. \& Chengalur, J.\ 2012, Experimental Astronomy, 33, 157. doi:10.1007/s10686-011-9279-5

\bibitem[Radhakrishnan \& Cooke(1969)]{1969ApL.....3..225R} Radhakrishnan, V. \& Cooke, D.~J.\ 1969, \aplett, 3, 225. 

\bibitem[Rathnasree \& Rankin(1995)]{1995ApJ...452..814R} Rathnasree, N. \& Rankin, J.~M.\ 1995, \apj, On the Approach to Stability of Pulsar Average Profiles, 452, 814. doi:10.1086/176349

\bibitem[Reddy et al.(2017)]{2017JAI.....641011R} Reddy, S.~H., Kudale, S., Gokhale, U., et al.\ 2017, Journal of Astronomical Instrumentation, 6, 1641011-336. doi:10.1142/S2251171716410117

\bibitem[Ruderman \& Sutherland(1975)]{1975ApJ...196...51R} Ruderman, M.~A. \& Sutherland, P.~G.\ 1975, \apj, 196, 51. doi:10.1086/153393

\bibitem[Scheuer(1968)]{1968Natur.218..920S} Scheuer, P.~A.~G.\ 1968, \nat, Amplitude Variations in Pulsed Radio Sources, 218, 5145, 920. doi:10.1038/218920a0

\bibitem[Sharma et al.(2025)]{2025ApJ...980...26S} Sharma, S.~S., Hashimoto, T., Goto, T., et al.\ 2025, \apj, Retracing the Cold Plasma Dispersion Law in Pulsar B0329+54: New Insights into Frequency-dependent Dispersion Measures, 980, 1, 26. doi:10.3847/1538-4357/ada956

\bibitem[Swarup et al.(1997)]{1997hsra.book..217S} Swarup, G., Ananthakrishnan, S., Subrahmanya, C.~R., et al.\ 1997, High-Sensitivity Radio Astronomy, 217

\bibitem[Tu et al.(2022)]{2022MNRAS.512.1906T} Tu, Z.~Y., Yuen, R., Wen, Z.~G., et al.\ 2022, \mnras, 512, 2, 1906. doi:10.1093/mnras/stac539

\bibitem[Yan et al.(2018)]{2018ApJ...856...55Y} Yan, Z., Shen, Z.-Q., Manchester, R.~N., et al.\ 2018, \apj, 856, 1, 55. doi:10.3847/1538-4357/aaae64


\end{thebibliography}
\end{document}